# Understanding the Characteristics of Frequent Users of Emergency Departments: What Role Do Medical Conditions Play?

Jens RAUCH[a,1], Jens HÜSERS[a], Birgit BABITSCH[b] and Ursula HÜBNER[a]
[a] *Health Informatics Research Group, Osnabrück University AS, Germany*
[b] *New Public Health, University Osnabrück, Germany*

**Abstract.** Frequent users of emergency departments (ED) pose a significant challenge to hospital emergency services. Despite a wealth of studies in this field, it is hardly understood, what medical conditions lead to frequent attendance. We examine (1) what ambulatory care sensitive conditions (ACSC) are linked to frequent use, (2) how frequent users can be clustered into subgroups with respect to their diagnoses, acuity and admittance, and (3) whether frequent use is related to higher acuity or admission rate. We identified several ACSC that highly increase the risk for heavy ED use, extracted four major diagnose subgroups and found no significant effect neither for acuity nor admission rate. Our study indicates that especially patients in need of (nursing) care form subgroups of frequent users, which implies that quality of care services might be crucial for tackling frequent use. Hospitals are advised to regularly analyze their ED data in the EHR to better align resources.

## 1. Introduction

Patient volume in hospital emergency departments (EDs) has increased steadily worldwide. Over the past decades the average annual growth rate in Germany was 4.9 % [1] with some German hospitals even reporting an increase of up to 9 % in 2015 [2]. This surge in demand is the major cause for ED overcrowding, which negatively impacts patient safety and mortality rates [3]. In light of this, considerable attention has been drawn to the group of patients who make use of ED services excessively.

Frequent users (FU) are conservatively defined as presenting at an ED for three or more times per year [4]. While FU account for 12 – 18 % of ED visits overall, their proportion to all ED patients amounts to merely 3.5 – 7.7 % [5]. The common belief, that frequent ED users present with light complaints and are unnecessarily clogging EDs, was not supported by most research. Although it was found, that FU are triaged more often with a lower urgency level than non-FU [6] and overestimate their need for emergency treatment [3], FU are often sick patients with chronic conditions and exhibit higher admission rates [4]. Yet FU do form a heterogeneous group [5]. Even though numerous studies have focused on FU related diagnoses, chronic conditions and socio-demographic features of FU, no studies so far have identified distinct subgroups of FU. An interesting finding, however, is that patients do not show patterns of frequent ED use over an extended period of time [4]. This suggests, that certain circumstances lead to frequent ED use that could be related to temporary issues of health or care. Especially, ambulatory care sensitive conditions (ACSC) [1] might pose a prominent risk factor, when ambulatory care fails to provide appropriate treatment or is out of reach. This study aims to further elucidate some of the constellations that elicit frequent use. We first examine how ACSC affect the risk of patients for exhibiting frequent use. Secondly, we determine distinct diagnoses subgroups of FU, regarding their morbidity, degree of acuity and admission rate. Finally, we readdress the issue, as to whether FU has an effect on complaint acuity or admission rate.

[1] Corresponding Author: Jens Rauch, Osnabrück University of AS, Health Informatics Research Group, PO Box 1940, 49009 Osnabrück, Germany; E-Mail: j.rauch@hs-osnabrueck.de.

## 2. Methods

We performed a retrospective longitudinal study including ED patients of an urban German teaching hospital (35,000 cases per year). Data were extracted from the hospital data warehouse (MS SQL Server 2016). Subjects were included only if they were at least 18 years old, triaged upon arrival and received at least one diagnosis. The resulting 23,364 patients corresponded to 29,520 ED visits. Patients were defined as FU, when they had three or more visits within the study period (January 1 until December 31, 2017). Complaint acuity was measured by triage (Manchester Triage System [7]). With most ICD-10 diagnosis codes occurring only once or rarely in our data set, we used the Clinical Classifications Software (CSS) [8] to collapse these into 260 meaningful categories. This led to a roughly uniform distribution of conditions. We used the complete set of 40 ACSC from Sundmacher et al. [9], since they comprise relevant conditions in an ED context [1].

For simple FU vs. non-FU group comparisons regarding demographic variables we performed a t-test (age) and chi square test (gender). In order to test, whether FU are associated with higher acuity and admission rates we performed multiple linear regression analyses for triage time and admission as criterion respectively with FU as predictor and adjusting for gender and age. Relative risk of frequent use for ACSC was determined by estimating FU prevalence per ACSC compared to prevalence among patients without ACSC. This was done (1) regarding main diagnoses only and (2) for all diagnoses, since ACSC might accompany the emergency condition in a systematic way.

For discovery of FU subgroups, we used nonnegative matrix factorization (NMF) as a means for clustering patients according to their CCS features, triage and admission. This procedure is well suited for identifying latent features in very sparse data [10], as is the case for the patients' diagnoses. To determine an optimal number of clusters, we performed 50 randomly initialized factorization runs for ranks in range 2 to 15. For each factorization, we calculated cophenetic correlation coefficients and residual of sum squares, by which optimal rank can be selected [11]. Cluster robustness was verified by computing consensus matrices. The resulting factors where then labeled semantically according to their associated diagnoses (CCS features).

## 3. Results

Frequent users made up 5.5 % (n = 1,292) of the study sample, accounting for about 15.5 % (n = 4, 583) of visits. Mean age in years for FU (64.1, SD = 21.0) and non-FU (54.0, SD = 21.8) differed significantly (p < .001), whereas gender ratio did not (53.1 % vs 52.7 % male; p = .81). Analyzing the risk of becoming a frequent user when having at least one of the ambulatory care sensitive conditions showed that a main diagnosis of heart failure increased the risk for frequent ED use by at least 75 %. Pressure ulcers, diseases of the urinary system as well as gastritis and duodenitis increased this risk at least by 25 % (Fig. 1 top). When including secondary diagnoses, risk for FU to be associated with certain ACSC was two to three times as high for patients with alcoholic liver disease, and at least 75 % higher for heart failure and malnutrition, bronchitis and COPD. ACSC with the least estimated risk for FU in both diagnose sets were soft tissue disorders and dental diseases. In accordance with Brunet [11], we defined k = 4 as optimal number of clusters for identifying disease subgroups of FU. Consensus matrices also showed a good cluster robustness for this k value. The four clusters with their respective CCS diagnosis codes, top associated ICD-10 codes and ACSC are shown in Table 1. Finally, regarding the potential influence of FU on admission rate and acuity, the regression models showed no significant effect of FU (yes/no) as variable of interest with respect to admission rate (p = .81), and a negligible effect of FU on acuity as measured by triage time (+1.2 minutes; p < .05).

**Table 1.** Diagnose clusters for frequent ED users

| Cluster | Summary | CCS codes | TOP ICD10 | TOP ACSC (main diagnoses) |
|---|---|---|---|---|
| I | Nutritional deficiencies, infections, hypertension, gastritis & duodenitis; triage: 30 min | 259, 3, 55, 159, 98 155, 52 | Z74, B96.2, E87, N39.0, I10 | Diseases of urinary system, intestinal infectious diseases |
| II | Device or implant complication, injuries, wounds, fractures; triage: 30 – 120 min; no admission | 237, 239, 235, 205 | T83, S30.0, S01, M54 | Mental and behavioral disorders due to use of alcohol or opioids, back pain |
| III | Immobility, dementia, incontinence; triage: unspecific | 95, 163, 68, 254 | R26, F03, R39, Z50 | Diseases of urinary system, metabolic disorders |
| IV | Heart failure, chronic renal failure, cardiac dysrhythmias; triage: 10 min | 108, 259, 158, 106 | I50.1, Z92.1, N18, I48 | Heart failure, ischemic heart diseases, bronchitis & COPD |

## 4. Discussion

This study contributes to a better understanding of what kind of patients typically present repeatedly at the ED of an urban teaching hospital. Analysis of FU in conjunction with ACSC implies, that certain conditions, such as alcoholic liver disease and heart failure, pose a strong risk for frequent ED visits. The results are in line with several other studies that have reported association to selected diagnoses or generally ICD-codes [4]. Our findings clarify the relation to ACSC by estimating relative risk for FU. Yet it remains to be determined whether these higher risks stem from individual circumstances or insufficient ambulatory care, especially for these specific conditions. However, many of the higher risk ACSC apparently have complex factors in common. Our approach specifically addresses the question of what diagnoses in FU typically go together and how they are related to ACSC. As could be shown, the results from FU cluster analysis outline four of these diagnose compounds. While both clusters I and III obviously relate to care dependent patients (as indicated by ICD-10 codes Z74 and R26), they comprise two clearly different groups of conditions. Interestingly, these clusters also relate to conditions, which are typical patients with nursing needs [12]. Cluster II is related to lighter ED cases, due to lower triage and ambulatory treatment and includes less distinct conditions. Cluster IV comprises patients with chronic heart and renal conditions that also go together with diabetes mellitus and have already been identified as typical for FU [13]. Clusters, in which ICD-codes and ACSC diverge, point to underlying causal mechanisms with varying manifestation (e.g. alcoholism, nursing care needs).

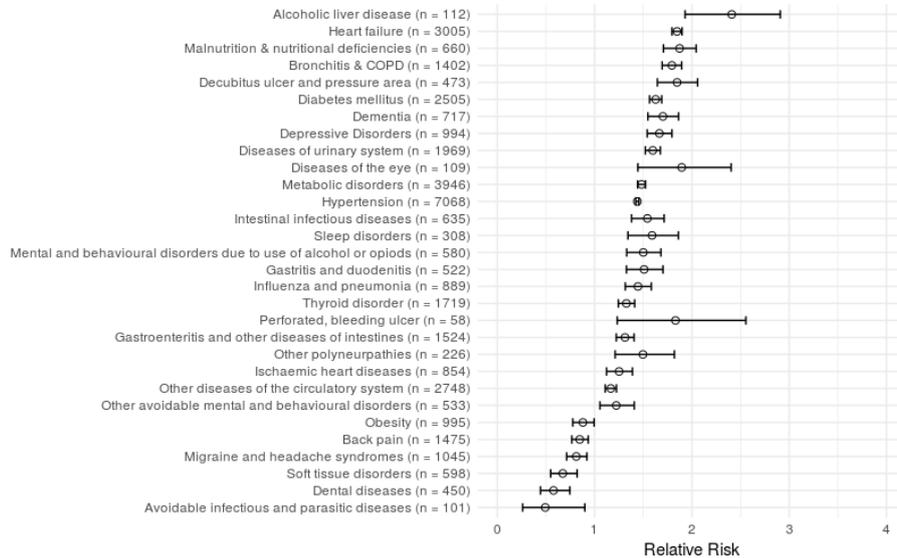

**Figure 1.** Relative risk of ED frequent use for main diagnoses (above) and all diagnoses (below), that qualify as ACSC, ordered by lower bound of confidence interval (95 %). Values express factors for risk change compared to non-FU (e.g. 1.5 means 50 % higher). Only conditions with at least 20 cases in overall data are shown.

The clusters presented here can serve as a starting point for characterizing FU subgroups in a more detailed fashion. Our results do not show any notable relation between FU and admission rates or acuity as compared to non-FU. Age however was highly significant and associated with both higher acuity and higher admission rates in our regression models. This might explain opposite findings that did not adjust for age (cf. [13]). This study has several limitations. Risk analysis for ACSC indicates that crucial factors are actually underlying causes. Alcoholism not being treated certainly leads to a number of conditions that end up in an ED sooner or later, no matter whether these qualify as ACSC or not. Ultimately, the extracted FU subgroups from factorization are too few in number to be isolated phenomena in the sense of patient phenotypes (cf. [14]). Therefore, subsequent clustering and phenotyping of FU groups needs to be pursued.

## 5. Conclusion

This is the first study to examine risk of frequent ED use for ACSC and to extract diagnose subgroups from frequent users. It contributes to a better understanding of common patterns and characteristics of patients, who excessively visit ED. Apparently, the most prominent FU subgroups are care dependent patients with typical ACSC, which points to quality of ambulatory care as a crucial factor that might explain frequent use. The study further demonstrates that secondary use of EHR data is highly desirable for hospitals. We recommend hospitals to regularly analyze their ED patients based on EHR data in this manner to better align resources and competencies in emergency departments.

## 6. Conflict of Interest

The authors declare no conflict of interest.

## Acknowledgements

This research is funded by the State of Lower Saxony, Project ROSE (ZN 3103). We would like to thank Dr. Mathias Denter, Jürgen Kleinschmidt, Ingo Mette, Daniel Spahn, Tobias Sonnenberg and Georg Schulte for their help and valuable input.